\documentclass[sigconf,screen]{acmart} 
\usepackage{microtype}
\usepackage{balance}
\usepackage{tabularx}
\newcommand{\tool}{\texttt{OwlEyes-Online}}

\AtBeginDocument{%
  \providecommand\BibTeX{{%
    \normalfont B\kern-0.5em{\scshape i\kern-0.25em b}\kern-0.8em\TeX}}}

\setcopyright{acmcopyright}
\acmPrice{15.00}
\acmDOI{10.1145/3468264.3473109}
\acmYear{2021}
\copyrightyear{2021}
\acmSubmissionID{fse21demo-p3-p}
\acmISBN{978-1-4503-8562-6/21/08}
\acmConference[ESEC/FSE '21]{Proceedings of the 29th ACM Joint European Software Engineering Conference and Symposium on the Foundations of Software Engineering}{August 23--28, 2021}{Athens, Greece}
\acmBooktitle{Proceedings of the 29th ACM Joint European Software Engineering Conference and Symposium on the Foundations of Software Engineering (ESEC/FSE '21), August 23--28, 2021, Athens, Greece}

\begin{document}

\title{OwlEyes-Online: A Fully Automated Platform for Detecting and Localizing UI Display Issues}

\author{Yuhui Su}
\authornote{Both authors contributed equally to this research.}
\authornote{Also With Laboratory for Internet Software Technologies, Beijing, China}
\affiliation{%
\institution{Institute of Software Chinese Academy of Sciences, University of Chinese Academy of Sciences}
\city{Beijing}
\country{China}
}

\author{Zhe Liu}
\authornotemark[1]
\authornotemark[2]
\affiliation{%
\institution{Institute of Software Chinese Academy of Sciences, University of Chinese Academy of Sciences}
\city{Beijing}
\country{China}
}

\author{Chunyang Chen}
\affiliation{%
  \institution{Monash University}
  \city{Melbourne}
  \country{Australia}
}

\author{Junjie Wang}
\authornotemark[2]
\authornote{Corresponding author}
\authornote{Also With State Key Laboratory of Computer Sciences, Beijing, China}
\affiliation{%
\institution{Institute of Software Chinese Academy of Sciences, University of Chinese Academy of Sciences}
\city{Beijing}
\country{China}
}

\author{Qing Wang}
\authornotemark[2]
\authornotemark[3]
\authornotemark[4]
\affiliation{%
\institution{Institute of Software Chinese Academy of Sciences, University of Chinese Academy of Sciences}
\city{Beijing}
\country{China}
}

\begin{abstract}
GUI provides visual bridges between software apps and end users.
However, due to the compatibility of software or hardware, UI display issues such as text overlap, blurred screen, image missing always occur during GUI rendering on different devices.
Because these UI display issues can be found directly by human eyes, in this paper, we implement an online UI display issue detection tool {\tool}, which provides a simple and easy-to-use platform for users to realize the automatic detection and localization of UI display issues.
The {\tool} can automatically run the app and get its screenshots and XML files, and then detect the existence of issues by analyzing the screenshots. In addition, {\tool} can also find the detailed area of the issue in the given screenshots to further remind developers. Finally, {\tool} will automatically generate test reports with UI display issues detected in app screenshots and send them to users.
It was evaluated and proved to be able to accurately detect UI display issues.
Tool Link: \url{http://www.owleyes.online:7476}.
Github Link: \url{https://github.com/franklinbill/owleyes}.
Demo Video Link: \url{https://youtu.be/002nHZBxtCY}.





\end{abstract}



\maketitle

\section{Introduction}
\label{sec_introduction}
GUI is widely used among modern mobile apps, making it practical and easy to use.
However, with the development of visual effects of GUI, five categories of UI display issues~\cite{liu2020owl} such as \textit{component occlusion, text overlap, missing image, null value} and \textit{blurred screen} always occur during the UI display process, especially on different mobile devices.
Detecting those issues is a hard problem because most of those UI display issues are caused by many factors, especially for Android, such as different Android OS versions, device models, and screen resolutions~\cite{wei2016taming}. 
Nowadays, some practical automated testing tools like Monkey~\cite{Monkey, Wetzlmaier2017Hybrid}, Dynodroid~\cite{Dynodroid} are also widely used in industry.
However, these automated tools can only spot critical crash bugs, rather than UI display issues that cannot be captured by common tools. Inspired by the fact that display bugs can be easily spotted by human eyes, we develop an automated online tool {\tool}\footnote{{\tool} is named as our approach is like the owl's eyes to effectively spot UI display issues. And our model (nocturnal like an owl) can complement conventional automated GUI testing (diurnal like an eagle) for ensuring the robustness of the UI.}, which provides quick detection and localization of UI display issues from apps or GUI screenshots.

\begin{figure*}[htb]
\vspace{-0.05in}
\centering
\includegraphics[width=16.5cm]{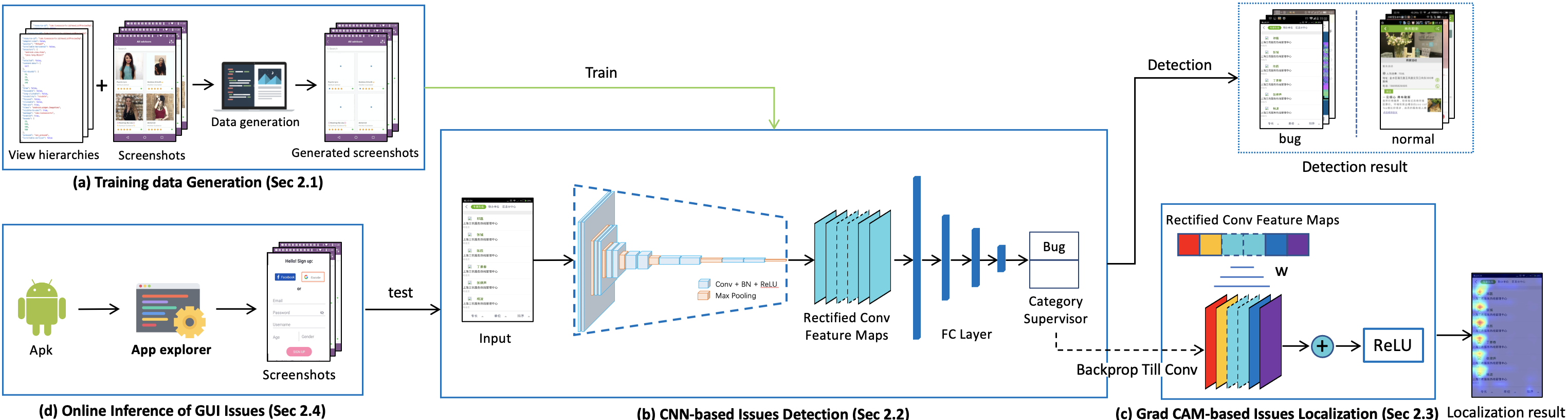}
\vspace{-0.05in}
\caption{Overview of {\tool}}
\label{fig:overview}
\vspace{-0.1in}
\end{figure*}

The {\tool} is a user-friendly web app. Developers can upload GUI screenshots or apps and receive accurate UI display issue detection results. When a developer uploads an APK, it will automatically run the app and get its screenshots, and then we use computer vision technologies to detect the UI display issues.
{\tool} builds on the CNN to identify the screenshots with issues and Grad-CAM to localize the regions with UI display issues in the screenshots for further reminding developers. 
Finally, it summarizes the detection and localization results, automatically generates the test report and sends it to users.
Considering that the CNN needs lots of training data, we adopt a heuristic data generation method to generate the training data.


{\tool} provides a dashboard for users to upload the screenshots or apps.
After analyzing an uploaded screenshot, it displays detection results in real-time. As for an app, it automatically generates a test report (issue screenshots, localization, etc.) and sends the report to the user in an email.

This paper makes the following contributions:
\begin{itemize}
\item We implement a CNN based issue detection method and a Grad-CAM based issue localization method to detect UI display issues from GUI screenshots.

\item We develop a fully automated web app. Users only need to upload an APK file, and {\tool} will automatically generate test reports and send them to users. We release the implementation of {\tool} on Github~\cite{GithubLink}.

\item An empirical study among professionals proves the value of our UI display issue detection method and {\tool}.

\end{itemize}

\section{Our Fully Automated Approach}
\label{sec_approach}
According to the features of UI display issues, we propose a fully automated UI display issue detection and localization approach. It mainly includes four parts, which are heuristic-based data generation, CNN-based issues detection, Grad CAM-based issues localization, and online inference of GUI issues.
As shown in Figure \ref{fig:overview}, to improve the accuracy of our model, we use the heuristic-based data generation method to generate a number of training data. Given an APK, {\tool} automatically runs it and collects screenshots. Then the CNN-based model classifies if they relate to any issues via the visual understanding. Once an issue is confirmed, our model can further localize its specific issue position on the UI screenshot by Grad CAM-based model to remind the developers. 

\begin{figure*}[htb]
\centering
\includegraphics[width=17cm]{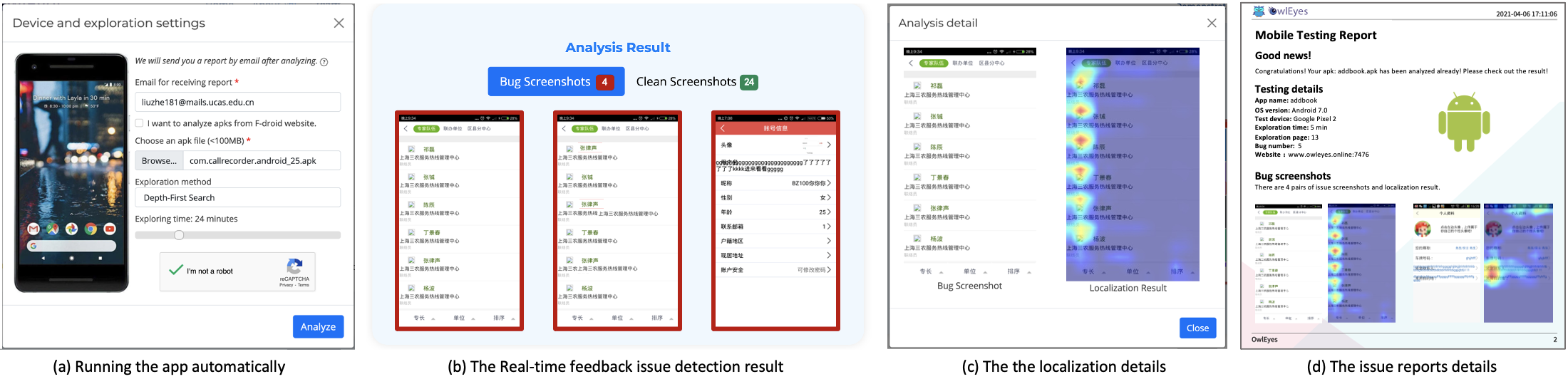}
\vspace{-0.05in}
\caption{Illustration of our {\tool} web application.}
\label{fig:Illustration}
\vspace{-0.1in}
\end{figure*}

\subsection{Heuristic-based Data Generation}
\label{subsec_approach_generation}
Training our proposed CNN for issues detection requires an abundance of screenshots~\cite{ResNet} with UI display issues.
However, there is so far no such type of open dataset, and collecting the related buggy screenshots is time- and effort-consuming.
Therefore, we develop a heuristic-based data generation method for generating UI screenshots with display issues from bug-free UI images in Figure \ref{fig:overview}(a).
The data generation is based on the Rico~\cite{Rico} dataset, which contains more than 66K unique screenshots and their JSON files (i.e., detailed run-time view hierarchy of the screenshot).
With the input screenshot and its associated JSON file, we first localize all the TextView and ImageView, then randomly chooses a TextView/ImageView depending on the augmented category.
Based on the coordinates and size of the TextView/ImageView, the algorithm then makes its copy and adjusts its location or size according to specific rules to generate the screenshots with corresponding UI display issues. 

\raggedbottom

\subsection{CNN-based Issues Detection}
\label{subsec_approach_CNN}
As the UI display issues can only be spotted via the visual information, we adopt the convolutional neural network (CNN)~\cite{lecun1998gradient,krizhevsky2012imagenet}, which has proven to be effective in image classification and recognition in computer vision~\cite{VGGNet,GoogleNet,ResNet}.
Figure \ref{fig:overview}(b) shows the structure of our model, which links the convolutional layers, batch normalization layers, pooling layers, and fully connected layers.
Given the input screenshot, we convert it into a certain image size with fixed width and height as the convolutional layer's parameters consist of a set of learnable filters. 
After the convolutional layers, the screenshots will be abstracted as a feature graph. 
In order to improve the performance and stability of CNN, we add Batch Normalization (BN)~\cite{BN} layers after the convolutional layer and standardize the input layer by adjusting and scaling activation. 
After the BN layer, we add the Rectified Linear Unit (ReLU) as the activation function of the network. 
The last several layers are fully connected neural networks (FC) which compile the data extracted by previous layers to form the final output. 
Finally, we obtain the detection results through softmax~\cite{PatternRecognition&ML}.

\raggedbottom

\subsection{Grad CAM-based Issues Localization}
\label{subsec_approach_localization}
As shown in Figure \ref{fig:overview}(c), we adopt the feature visualization method to localize the detailed position of the issues to remind the developers.
We apply the Grad-CAM model for the localization of UI display issues.
Gradient weighted Class Activation Mapping (Grad-CAM) is a technique for visualizing the regions of input that are ``important'' for predictions on CNN-based models~\cite{Grad-CAM} .
First, a screenshot with the UI display issue is input into the trained CNN model, and the category supervisor to which the image belongs is set to 1, while the rest is 0. 
Then the information is propagated back to the convolutional feature map of interest to obtain the Grad-CAM positioning. 
Through the feedback of global average pooling of the gradient, the weight $ \alpha_k^b $ of the importance of neurons is obtained. This weight captures the importance of the feature map $K$ of the target category $ b $ (Bug). By performing the weighted combination of the forward activation graph, we can obtain the class-discriminative localization map.
Finally, the point multiplication with the backpropagation can obtain the Grad-CAM as the result of issue localization. 

\subsection{Online Inference of GUI Issues}
\label{subsec_approach_Collection}

We use 20,000 screenshots generated in section \ref{subsec_approach_generation} to train our issue detection and localization model. Before the issue detection, we need to preprocess the APK submitted by the user online.
As shown in Figure \ref{fig:overview}(d), the user provides an Android APK, and we use the dynamic analysis method to run the app automatically to obtain the screenshots.
In detail, by leveraging the idea of dynamic app GUI testing~\cite{li2017droidbot,cai2020fastbot,Monkey,su2017guided}, we adopt an app explorer~\cite{li2017droidbot} to automatically explore the pages within an application through interacting with apps using random actions, e.g., clicking, scrolling, and filling in text. 
We also provide three testing strategies for users to choose from: Depth-First-Search (DFS)~\cite{shwail2013probabilistic}, Breadth-First-Search (BFS)~\cite{beamer2012direction}, and random exploration~\cite{Monkey}.

\section{Tool Implementation And Usage}
\label{sec_implementation}

{\tool} is a web app, which provides a convenient tool for users to detect and localize the UI display issues in the GUI screenshots.

\subsection{Web Implementation}
\label{subsec_Web}
{\tool} can automatically run applications and generate test reports for users. We customized the deep learning model in PyTorch. The {\tool} consists of two parts: running the application automatically, feeding back the test results in real-time.

\textbf{Running the app automatically:} Figure \ref{fig:Illustration}(a) shows an example of our running the app automatically page. Users can upload the APK or its download link. In addition, we allow users to customize the exploration strategy, select the appropriate device, and some personalization settings to provide a more friendly interactive experience.

\textbf{The Real-time feedback issue detection results:} This page in Figure \ref{fig:Illustration}(b) will give real-time feedback test results while running the application automatically. On this page, we implement some functions to provide a more friendly interactive experience, including:

\textbf{Click to view the localization details:} In Figure \ref{fig:Illustration}(c), click the screenshot of the UI display issue to view the localization of it in the screenshot (in the form of a thermal graph).

\textbf{Export test report}: In Figure \ref{fig:Illustration}(d), users fill in e-mail information, and we will automatically generate test reports and send them to users. The test report includes the number of issues of the application and the screenshots of the issue and the XML corresponding to the screenshots.

\raggedbottom

\subsection{Model Implementation}
\label{subsec_Model}
Our CNN model is composed of 12 convolutional layers with batch normalization, 6 pooling layers, and 4 full connection layers for classifying UI screenshot with display issues. 
The size of a convolutional kernel in the convolutional layer is 3 $\times$ 3. We set up the number of convolutional kernels as 16 for convolutional layer 1-4, 32 for convolutional layer 5-6, 64 for convolutional layer 7-8, and 128 for convolutional layer 9-12. 
For the pooling layers, we use the most common-used max-pooling settings~\cite{simard2003best}, i.e., pooling units of size 2 $\times$ 2 applied with a stride ~\cite{VGGNet}. 
We set the number of neurons in each of the fully connected layers as 4096, 1024, 128, and 2 respectively.
For data preprocessing, we rotate some UI of the horizontal screens to vertical, and resize the screens to 768 $\times$ 448. 
We implement our model based on the PyTorch~\cite{pytorch} framework. 
The model is trained in an NVIDIA GeForce RTX 2060 GPU (16G memory) with 100 epochs for about 8 hours.

\subsection{Usage Scenarios}
\label{subsec_Usage}

We present several examples to illustrate how developers would interact with {\tool}. In some cases, developers collect a large number of screenshots of applications (such as crowdtesting platform, automatic testing). However, these automated tools can only spot critical crash bugs, rather than UI display issues that cannot be captured by common tools. Developers can upload application screenshots to our {\tool} directly. {\tool} will analyze the screenshots and detect the UI display issue in the screenshots.

For testing whether UI display issues exist in the application, developers can directly upload an APK to our {\tool} , which will automatically explore the application and detect UI display issues. Developers can also customize the exploration method and duration and submit the e-mail information. After the issue detection, {\tool} will automatically generate the issue report and send it to the developer's e-mail. Considering the network delay, developers can also upload an application's download link, and {\tool} will automatically download the APK in the background for testing.
\section{Evaluation}
\label{sec_evaluation}
The goal of our study is to evaluate the usefulness of our platform {\tool} in terms of (i) its effectiveness in detecting and localizing UI display issues, and (ii) the usability of our {\tool}.

\subsection{Effectiveness Measurement}
\label{sec_effectiveness}

Given the effectiveness of our {\tool} for UI display issue detection, we conduct experiments on 8K Android mobile GUI collected by one of the largest crowd-testing platforms~\cite{BaiduC}.
This part is also published in our previous work~\cite{liu2020owl} and and we mainly use evaluation metrics of precision and recall.

Table \ref{tab:RQ2-Baseline} shows the performance comparison with the baselines. 
With {\tool}, the precision is 0.85 and the recall is 0.84.
We can see that our proposed {\tool} is much better than the baselines, i.e., 58\% higher in precision and 17\% higher in recall compared with the best baseline, Multilayer Perceptron (MLP).
This further indicates the effectiveness of {\tool}.
Besides, it also implies that {\tool} is especially good at hunting for the buggy screenshots from candidate ones, i.e., significant improvement in recall.

\begin{table}[H]
\vspace{-0.1in}
\caption{Performance comparison with baselines}
\vspace{-0.1in}
\label{tab:RQ2-Baseline}
\centering
\footnotesize
\begin{tabular}{p{1.7cm}<{\centering} | p{1.5cm}<{\centering} | p{1.5cm}<{\centering} | p{1.5cm}<{\centering}}
\hline
\textbf{Method} & \textbf{Precision} & \textbf{Recall} & \textbf{F1-score} \\
\hline
RF-SIFT & 0.458 & 0.458 & 0.432 \\ 
RF-SURF & 0.513 & 0.524 & 0.519 \\ 
RF-ORB & 0.520 & 0.528 & 0.524 \\  
\hline
MLP & 0.537 & 0.727 & 0.618 \\  
\hline
\textbf{{\tool}} & \textbf{0.850} & \textbf{0.848} & \textbf{0.849} \\ 
\hline
\end{tabular}
\vspace{-0.1in}
\end{table}


\subsection{Usefulness Measurement}
\label{sec_usefulness}

To further assess the usefulness of our approach, we randomly sample 2,000 Android applications from F-Droid~\cite{F-droid} and 1,000 applications from Google Play~\cite{GooglePlay}.
Note that none of these apps appears in our training dataset.
Among the 3,000 collected applications, 59\% (1756/3000) applications can be successfully run with {\tool}.
For the remaining 1,756 applications, an average of 8 screenshots is obtained for each application. 
We then feed those screenshots to our {\tool} and detect if there are any display issues.
Once a display issue is spotted, we create a bug report by describing the issue attached with a buggy UI screenshot.
Finally, we report them to the app development team through issue reports or emails. 
Our {\tool} has detected 113 UI display issues, among which 35 have been confirmed and 29 have been fixed.
These fixed or confirmed bug reports further demonstrate the effectiveness and usefulness of our proposed approach in detecting UI display issues. 

Regarding the user experience of our {\tool}, we create an online survey on 20 professional developers, testers, and researchers, all of whom major in computer science with more than 3 years of app testing or developing experience. 10 of them are from the industry with practical working experience\footnote{Some testers are from NVIDIA, Citibank, Sony, Baidu, Alibaba, Three Fast Online, and ByteDance.}. We ask them to use our {\tool} and ask them about the usefulness of the {\tool} for their work, as well as its potential and scalability in the future. In the end, participants fill in the System Usability Scale (SUS) questionnaire~\cite{brooke1996sus} (5-point Likert scale~\cite{5-likert-1} from 1 (strongly disagree) to 5 (strongly agree)). The questionnaire also asks participants to select the TechLand system features that they deem most useful or least useful for the tasks.

\begin{figure}[htb]
\centering
\includegraphics[width=7.3cm]{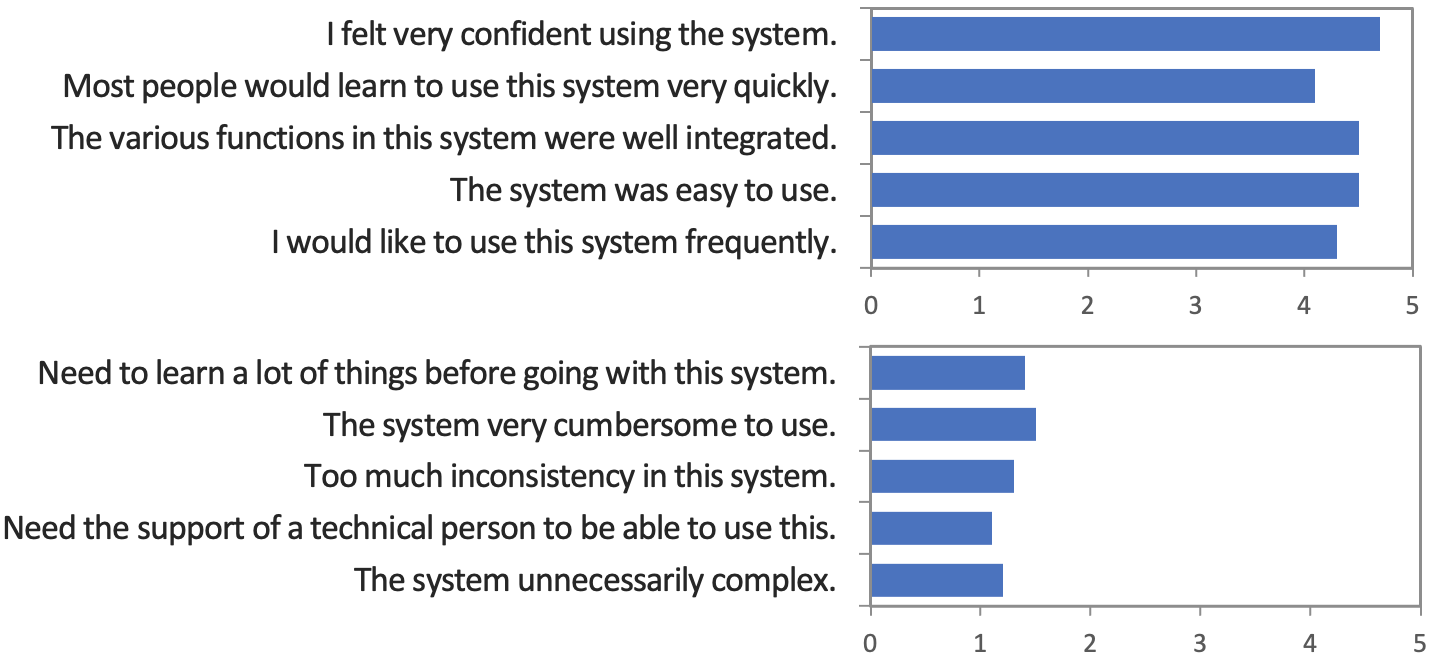}
\vspace{-0.1in}
\caption{Average score of SUS results}
\label{fig:usefulness}
\vspace{-0.15in}
\end{figure}

Figure \ref{fig:usefulness} summarizes the participants’ ratings of the 10 system design and usability questions in the System Usability Scale questionnaire. The upper half of figure \ref{fig:usefulness} shows that participants agree or strongly agree that our system is easy to use and the features of the {\tool} system are well-devised. The lower half of figure \ref{fig:usefulness} further confirms the simplicity and consistency of our {\tool} system. Furthermore, the average helpfulness of the {\tool} system for the tasks is 4.42, which indicates that participants appreciate the help of the {\tool} system in the tasks.
All participants indicated that {\tool} has a good UI display issue detection effect. Among these professionals, 10 of them are working on app testing. They think {\tool} can help them localize the UI display issues more quickly. 7 Android developers said that our issue localization model helps them better localize the issue on the UI interface so that they can better repair the issue later. Among them, 4 developers hope we can further give the possible repair methods and causes of these issues. The other 3 participants who are studying GUI testing also indicated that they hope we can analyze the cause of issue in the next stage. They think that using the visual information of application screenshots is a very helpful and engaging work.
\section{Conclusion}
\label{sec_conclusion}

Improving the quality of mobile applications, especially in a proactive way, is of great value and always encouraged. 
In this demo, we show {\tool}, a fully automated UI display issue detection and localization tool. We use dynamic analysis to explore the application automatically and get its screenshots. And users can customize the exploration time, exploration strategy and so on. Then we can complete the detection and localization of UI display issues based on CNN and Grad-CAM. Finally, we automatically generate test reports and send them to users. 
We evaluate it from two aspects of detection accuracy and tool practicability.
The {\tool} is proven to be effective in real-world practice, i.e., 64 confirmed or fixed previously undetected UI display issues from popular Android apps.
It also achieves boosts of more than 17\% and 23\% in recall and precision compared with the best baseline.
The evaluation shows that {\tool} is a good starting point for UI display issue detection.

In the future, we will further study the root cause of UI display issue.
Finally, according to the issue category, we will devise a set of tools for recommending patches to developers to fix the UI display issues.


\section*{ACKNOWLEDGMENTS}
This work is supported by the National Key Research and Development Program of China under grant No.2018YFB1403400, National Natural Science Foundation
of China under Grant No. 62072442, No. 62002348.

\balance
\bibliographystyle{ACM-Reference-Format}
\bibliography{reference}

\end{document}